\documentclass[10pt]{bmc_article}    

\usepackage{cite} 
\usepackage{hyperref}  
\usepackage{ifthen}  
\usepackage{multicol}   
\usepackage[utf8]{inputenc} 
\usepackage{tikz}
\usetikzlibrary{positioning,arrows.meta,arrows,backgrounds,patterns}

\usepackage{graphicx}

\newboolean{publ}

\newenvironment{bmcformat}{\fussy\setboolean{publ}{true}}{\fussy}

\begin{document}
\begin{bmcformat}

\title{Automatic quantification of the microvascular density on whole 
slide images, applied to paediatric brain tumours}


\author{Christophe Deroulers\correspondingauthor$^1$%
       \email{Christophe Deroulers\correspondingauthor - deroulers@imnc.in2p3.fr}%
      \and
         Volodia Dangouloff-Ros$^{2,3}$%
         \email{Volodia Dangouloff-Ros - volodia.dangouloff-ros@aphp.fr}
      \and
         Mathilde Badoual$^1$%
         \email{Mathilde Badoual - badoual@imnc.in2p3.fr}
      \and
         Pascale Varlet$^{3,4,5}$%
         \email{Pascale Varlet - varlet@ch-sainte-anne.fr}
       and 
         Nathalie Boddaert$^{2,3,4,6}$%
         \email{Nathalie Boddaert - nathalie.boddaert@aphp.fr}%
      }
      

\address{%
    \iid(1)Univ Paris Diderot, Laboratoire IMNC, UMR 8165 CNRS, %
        Univ Paris-Sud, F-91405 Orsay, France\\
    \iid(2)Department of Paediatric Radiology, H\^opital Necker Enfants%
        Malades, AP-HP, 75105 Paris, France\\
    \iid(3)INSERM U1000, Paris, France\\
    \iid(4)Univ Paris Descartes, Paris, France\\
    \iid(5)Department of Neuropathology, Centre Hospitalier Sainte-Anne, %
        Paris, France
    \iid(6)UMR 1163, Institut Imagine, Paris, France
}%

\maketitle


\begin{abstract}
        \paragraph*{Background:}

Angiogenesis is a key phenomenon for tumour progression, diagnosis and 
treatment in brain tumours and more generally in oncology. Presently, 
its precise, direct quantitative assessment can only be done on whole 
tissue sections immunostained to reveal vascular endothelial cells. But 
this is a tremendous task for the pathologist and a challenge for the 
computer since digitised whole tissue sections, whole slide images 
(WSI), contain typically around ten gigapixels.
      
        \paragraph*{Methods:}

We define and implement an algorithm that determines automatically, on a 
WSI at objective magnification $40\times$, the regions of tissue, the 
regions without blur and the regions of large puddles of red blood 
cells, and constructs the mask of blur-free, significant tissue on the 
WSI.

 Then it calibrates automatically the optical density ratios of the 
immunostaining of the vessel walls and of the counterstaining, performs 
a colour deconvolution inside the regions of blur-free tissue, and finds 
the vessel walls inside these regions by selecting, on the image 
resulting from the colour deconvolution, zones which satisfy a 
double-threshold criterion. The two thresholds involved are 
automatically computed from the WSI so as to cope with variations in 
staining and digitisation parameters. A mask of vessel wall regions on 
the WSI is produced.

 The density of microvessels is finally computed as the fraction of the 
area of significant tissue which is occupied by vessel walls.

We apply this algorithm to a set of 186 WSI of paediatric brain tumours 
from World Health Organisation grades I to IV.

        \paragraph*{Results:}

 The algorithm and its implementation are able to distinguish on the WSI 
the significant tissue and the vessel walls. The segmentations are of 
very good quality although the set of slides is very heterogeneous (in 
tumour type, in staining and digitisation parameters, and inside WSI 
themselves, where the tissue was often very fragmented). The computation 
time is of the order of a fraction of an hour for each WSI even though a 
modest desktop computer is used (a 2012 Mac mini) and the average size 
of WSI is 7~gigapixels. The computed microvascular density is found to 
be robust. We find that it strongly correlates with the tumour grade.

        \paragraph*{Conclusions:}

 We have introduced a method of automatic segmentation of significant, 
blur-free tissue and of vessel walls, and of quantification of the 
density of microvessels, in WSI. We successfully tested it on a large 
variety of brain tumour tissue samples. This method requires no training 
and estimates automatically several important parameters of the 
segmentation. It is robust and can easily be applied to other tumour 
types and other stainings. It should improve the reproducibility of 
quantitative estimates in pathology while sparing the pathologist time 
and effort.

\end{abstract}

\ifthenelse{\boolean{publ}}{\begin{multicols}{2}}{}


\section*{Introduction}

 Angiogenesis is one of the key features of tumour progression, 
sustaining growth and sometimes enabling a change of aggressiveness when 
it starts~\cite{angiogenese}. In brain 
tumours~\cite{angiogenese-et-invasion-gliomes, 
angiogenese-gliomes-malins}, it is a crucial histology criterion used 
in diagnosis and to classify the disease into the proper World Health 
Organisation (WHO) grade~\cite{classification-oms-snc-2016}. Therefore, 
it is of great importance to be able to quantify in a reliable and 
robust way the status of the tumour vascular system. Although there 
exist several noninvasive, macroscopic imaging 
techniques~\cite{mesure-angiogenese-colorectal-perfusionct, 
imagerie-rm-perfusion, 
imagerie-fonctionelle-tumeurs-cerebrales-adultes-enfants}, not all of 
them are innocuous (they may use ionising radiations or contrast 
agents), and they don't yield a direct access to the geometric 
parameters of the vasculature. In contrast, after proper 
immunohistochemical staining~\cite{cd31, cd34}, biopsy samples reveal 
directly the tumour microvessels. It has been shown that the 
microvascular density, as measured on histology sections, is of 
prognostic significance in several brain 
tumours~\cite{angiogenese-valeur-pronostique-pour-tumeurs-cerebrales, 
volodia-asl}.

 However, assessing manually the density of microvessels on whole 
histology sections is a tremendous task, very hard to perform for a 
human, and prone to much inter- (and even intra-) individual variability 
and lack of reproducibility. Quantifying the vascularity only on a few 
randomly chosen regions, or on a few ``representative'' regions on the 
section~\cite{kayser2003} will make the task easier (shorter) for the 
pathologist, but will increase the measurement variability and might 
reinforce the subjectivity of the task.

 Luckily, virtual microscopy and the digitisation of pathology slides 
have become quite common over the last few years, allowing the use of 
the computer to perform various quantitative and reproducible 
measurements on histology 
sections~\cite{revue-analyse-image-histopathologique}. In the beginning 
of this digital era, for cost and material reasons, it was not possible 
to measure the parameters of microvessels at full resolution (20x or 
40x), and it was suggested to use images at resolution around 
1x~\cite{quantification-vaisseaux-caen, 
choix-resolution-pour-quantification-vaisseaux}. However, slide scanners 
produce now high resolution microscopy images of whole slides in a short 
time (at most a few minutes) and for a reasonable 
price~\cite{virtual-microscopy}, and using them for quantification 
should improve the precision of the results. The drawback of this high 
resolution is the very large size of the resulting files, which require 
specific software tools to be managed, like the ones some of us have 
already developed~\cite{ndpitools-article, largetifftools, ndpitools}.

 Such a quantification of microvessels on high-resolution images has 
already been undertaken by several groups~\cite{caiman, 
vascularity-from-traveling-salesman-distance, angiopath}. However, most 
of them were limited to small excerpts of the whole slide images (WSI), 
and possibly to relatively homogeneous sets of slides. Here, we report 
on a set of techniques we have developed to assess the density of 
microvessels on WSI of sizes a few tens of gigapixels, immunostained 
with CD34 (to reveal vascular endothelial cells), within a few minutes 
minutes, in a robust way, with a careful determination of zones of 
tissue without blur, and with as little intervention of the pathologist 
as possible. In particular, no training of an algorithm is necessary, 
hence the time-consuming task of manual segmentation of a number of 
vessels to feed to the computer is spared.

 Although immunofluorescence is able to provide valuable additional 
quantitative information about the vasculature, such as 3D 
aspects~\cite{quantification-vaisseaux-immunofluorescence}, it requires 
the use of more elaborate microscopy and it not yet compatible with 
clinical routine. Therefore, we stick to classical bright field 
microscopy of immunostained pathology thin sections.

 To demonstrate the versatility and scalability of our method, we 
applied it to a series of 129 human patients (186 WSI in total) 
suffering brain tumours of 19 different combinations of histological 
type and location, ranging from WHO grade I to grade IV. Since WSI are 
very large (our largest image had $162688\times98816$ pixels), they 
can't be opened in full in a standard computer's memory (we would need 
up to 60 GiB of RAM), and we had to develop strategies to treat them 
entirely without restricting ourselves to small excerpts. We used only 
open source software, or software we developed based on existing open 
source libraries, to avoid black-box algorithms, to promote 
interoperability and reproducibility, to reduce costs, and to avoid 
conflicts of interest~\cite{openaccess-2}.

\section*{Material and Methods}

 Our goal is to quantify the density of microvessels on a WSI as the 
ratio of the area occupied by vascular endothelial cells to the area 
occupied by the tissue. We assume that the WSI was obtained as a 
(possibly pyramidal) tiled TIFF or 
BigTIFF~\cite{bigtiff-specification-libtiff} file by digitisation at 
objective magnification $20\times$ or $40\times$ of a 5$\mu$m-thick 
tissue section of formalin-fixed, paraffin-embedded tissue, and that 
immunostaining with a CD34 antibody was performed so that microvessels 
appear brown whereas cell nuclei appear blue. The total area of tissue 
is typically of a few square centimetres.

 If some deviation from this protocol is in order, it should be easy to 
adapt our method. E.g., if the WSI is stored in another format, it can 
be converted to TIFF using the free software 
NDPITools~\cite{ndpitools,ndpitools-article} that we developed or 
OpenSlide~\cite{openslide1, openslide2}. If the colours after 
immunostaining are different, two parameters can be changed (see below).

\subsection*{Method overview}

 The aim is first to select the zone of tissue on the WSI, then the zone 
of vessel walls inside the zone of tissue, and finally to measure the 
areas of the two and compute their ratio. Because of unavoidable 
slide-to-slide variations in staining and digitisation parameters (e.g. 
light intensity or temperature colour), both zone selections will require 
prior calibration steps. And, due to technical details, the work flow 
will be slightly more complex. We must: \begin{itemize} \item exclude 
from the WSI regions where the image is not sharp enough to recognise 
vessel walls accurately

\item exclude from the WSI regions which look like tissue but should not 
be counted as such: essentially large puddles of red blood cells, 
coverslip boundaries and dust

\item not count as vessel walls extra-vascular CD34-positive tumour 
cells.
\end{itemize}

 A scheme of the whole process is shown in Figure~1.

\subsection*{Preparatory steps}

 From the full-resolution $40\times$ image, a $20\times$ image was 
generated by bilinear interpolation and stored into a 
JPEG-compressed~\cite{jpeg} tiled TIFF~\cite{tiff} file. Indeed, such an 
image proved of sufficient quality for several of the steps below while 
saving computation time.

 Then, a mosaic of the $20\times$ image was made and stored into JPEG 
files. This is a decomposition of the original image in rectangular 
pieces of equal sizes, stored into independent files for easy 
independent treatment, such that the original image is recovered if the 
pieces are reassembled together. We requested that each piece need at 
most 128~MiB to be stored (uncompressed) in RAM and that the dimensions 
of each piece be multiples of 8~pixels. This can be easily achieved 
using the \texttt{-m} and \texttt{-M} options of the 
\texttt{tiffmakemosaic} software~\cite{ndpitools,ndpitools-article}.

\subsection*{Selection of sufficiently sharp zones}

 We used a variation of the method of blur quantification 
of~\cite{stack-or-trash, these-david-ameisen}. On each piece of the 
$20\times$ mosaic, after decompression into RGB colour space, we applied 
a colour space transformation into HSV colour space as defined by the 
\texttt{vips} program~\cite{vips}, extracted the V channel and 
convoluted the resulting image with the Laplacian kernel $$\left( 
\begin{tabular}{ccc} -1 & -1 & -1 \\ -1 & 8 & -1 \\ -1 & -1 & -1 
\end{tabular} \right).$$

 Then, considering the result of the convolution as a mosaic of blocks 
of $8 \times 8$ pixels, we computed the $S_2$ 
score~\cite{these-david-ameisen}, namely the ratio of the sum of pixel 
intensities which are at least 10\% of the maximum intensity in the 
block to the sum of all 64~pixel intensities (or 0 if the denominator is 
vanishing). Such a division into blocks of $8 \times 8$ pixels is 
natural since it is at the basis of the JPEG compression used by most 
slide scanners and used by our method to store the $20\times$ image 
while saving disk space; artefacts due to this blocking are already 
present in the original WSI.

 We generated a graylevel image where the $S_2$ value of each block of 
the $20\times$ image was encoded as the intensity of one pixel (between 
0 and 255 included). The resulting image, which can be deemed a 
sharpness map, is 16 times smaller (in linear dimension) than the 
original $40\times$ image, thus has 256 times less pixels, and could 
easily be stored in a single file and opened at once in the computer's 
RAM.

 Finally, this $2.5\times$ sharpness map was transformed (using the 
ImageJ software~\cite{imagej, imagej-nature-methods}) into a mask of 
sharp regions in the following way: pixel intensities were averaged over 
regions of radius~2~pixels; a mask of regions where the resulting 
intensity is 43 and above was created (these are the sharp regions); 
sharp regions of less than 20~pixels of area (at resolution $2.5\times$) 
were turned into blurred regions, then blurred regions of less than 
20~pixels of area were turned into sharp regions. The final mask of 
sharp regions was written on the disk.

\subsection*{Selection of tissue zones}

 We used as a first criterion to distinguish tissue from background the 
value of brightness of pixels (B in HSB colour space as defined by 
ImageJ~\cite{imagej}). Therefore, we needed to calibrate the brightness of 
pixels in the background.

 This calibration may be influenced by the sides of the coverslip (and 
zones beyond) which are visible on 10 of our WSI. For each of these few 
images, we manually contoured the side of the coverslip on a downscaled 
$0.625\times$ image and stored the resulting contour (union of polygons) 
as ImageJ's \texttt{roi} files which will be subsequently read at the 
proper stage.

 We generated a $2.5\times$ image from the full resolution image. With 
ImageJ, we selected on this image the non-excluded zones (sides of the 
coverslip), then we extracted the brightness of pixels (between 0 and 
255) and selected pixels the brightness of which differed by less than 
$\pm 1.2$ from the Gaussian-weighted average (with standard deviation 
$\sigma=0.005$) over their vicinity. We selected connected regions of at 
least 160~pixels among these ``uniform'' pixels and took the 
intersection with regions where the brightness was 127 or above. This 
defined the reference regions for the background.

 We measured the histogram of the brightness value of the pixels in the 
reference regions. It always exhibits a peak of occurrence numbers of 
levels between 200 and 255. We measured the right end $b_{bg,r}$ of this 
peak, as the largest brightness level which occurs at least once. Then 
we measured the left end $b_{bg,l}$ of this peak using the following 
algorithm: starting from the right end, we scanned occurrence numbers of 
decreasing brightness levels. We stopped when the current occurrence 
number was below half of the largest observed occurrence number so far 
and either was zero or was larger than the last seen occurrence number.

 These left- and right-end define rather accurately the brightness of 
pixels belonging to the background and we stored them in a text file for 
later reuse. They had to be determined for each WSI because of marked 
variability: the left-end ranged from 212 to 234 while the right-end 
ranged from 232 to 249.

 In addition, we selected pixels within the reference regions which had 
a brightness between the left-end of the peak and the left-end plus 
three (included) and computed their average values of R, G and B 
(hereafter called $R_{bg}$, $G_{bg}$, $B_{bg}$). We stored these values 
in a text file for later use.

 The actual selection of tissue zones was performed on pieces of a 
mosaic of the image at resolution $20\times$. We produced with 
\texttt{tiffmakemosaic} a mosaic such that no piece required more than 
100~MiB of RAM to be opened, with overlap of 256~pixels between adjacent 
pieces. Pieces were stored as TIFF files with zip compression (rather 
than JPEG compression to avoid another information loss and to 
facilitate opening by ImageJ). On each piece, ImageJ was used to produce 
a mask of the tissue zones in the following way. We selected the pixels 
the brightness of which was outside the interval measured previously as 
``background intensity peak'', $[b_{bg,l}, b_{bg,r}]$ --- call them 
A-pixels. Then, among the connected regions formed by these pixels, we 
selected those which contained at least one pixel the saturation of 
which (in HSB space) was larger than the average of the saturation of 
all A-pixels --- thus constructing the B-regions. This was to prevent 
selecting uniform regions with unusual high or low brightness, which 
could be a large piece of dust or a pen stroke (uniformly black region).

 The resulting regions could contain ``holes'', some of which were 
regions of background inside a region of tissue. We noticed that almost 
all holes of less than $\approx$ 10000~pixels (at $20\times$ 
magnification) had to be considered as tissue. Therefore, we included in 
the B-regions the holes of area less than 10000~pixels and which did not 
touch the boundaries of the mosaic piece. Indeed, a hole touching a 
boundary either was a small part of a bigger hole (more than 
10000~pixels) extending on another mosaic piece, which should not be 
restored into the B-regions, or a small part of a small hole which would 
quite probably be entirely included into an adjacent mosaic piece, 
because mosaic pieces overlapped by 256 pixels. The exception would 
concern only small holes (less than 10000~pixels), a dimension of which 
exceeded 256 pixels, that is rather elongated holes, which was rather 
uncommon.

 Similarly, we found that holes having a fractal-like shape should also 
be restored as tissue in the B-regions. They were characterised in the 
following way: after we performed an \texttt{erode} then a 
\texttt{dilate} operation on them, their area was less than 
10000~pixels, and they did not touch the boundary of the mosaic piece.

 Then we made the boundaries of the B-regions (tissue regions) more 
regular applying an \texttt{erode} then a \texttt{dilate} on their mask. 
This eliminated the small overhangs or invaginations of a few pixels.

 We eliminated from the B-regions the connected regions of less than 
20000~pixels which were at a distance 30 pixels or larger from a region 
or 20000~pixels or more. Indeed, most of those were found to be non 
significant (dust, isolated cells, small pieces of tissue torn apart). 
And in the end we are interested only in the fraction of tissue area 
covered by vessels, which we neither overestimate nor underestimate by 
mistakenly removing a small significant area of tissue. This last 
operation was performed in ImageJ by a combination of morphological 
operations, including thresholding at 15 the distance map from the 
B-regions.

 Finally, we found that B-regions with too low saturation (greyish 
regions) could exist and should be disregarded as tissue (tissue 
contains at least blue cell nuclei or brown vessel walls). Therefore, 
after convoluting by a Gaussian kernel of standard deviation 
$\sigma=3$~pixels the image of the saturation in HSB colour space, we 
applied the isodata algorithm~\cite{isodatathreshold} to find an 
automatic threshold on this image and we defined as high-saturation 
pixels the pixels the saturation of which was above the isodata 
threshold and above 30. Only B-regions containing at least one 
high-saturation pixel were kept as tissue regions.

 The mask (binary image) of the tissue regions of each mosaic piece was 
saved as PNG files, then all PNG files were combined into a large, 
$20\times$ resolution, mask of tissue thanks to an in-house developed C 
program for the sake of speed and RAM economy. The whole mask was stored 
on the disk as a 1-bit-per-pixel Deflate-compressed~\cite{tiff} TIFF 
file, which achieves a very high level of compression. Then the pieces 
of the mosaic with overlap were erased from the disk.

 If there existed, for this WSI, a \texttt{.roi} file defining the 
side(s) of the coverslip on the image, the mask of the region to exclude 
was formed as a binary image of resolution~$0.625\times$ and the mask of 
tissue was replaced by the result of a logical \texttt{and} of the 
former mask of the tissue and the inverse of the mask of the region to 
exclude blown up to resolution~$20\times$. This logical operation was 
performed thanks to another in-house developed C program, again for the 
sake of speed and RAM economy.

 Finally, the logical \texttt{and} of the mask of sharp regions and of 
the mask of tissue was computed through the same C program and stored on 
the disk as a bilevel, Deflate-compressed TIFF file. Again, although the 
resulting mask is a very large image (1,76~gigapixels in average), this 
form of compression is very efficient: the average disk size of the mask 
was 4,36~MiB, ranging from 0,26~MiB to 16,38~MiB.

 A detail from an example of such a mask of sharp tissue is shown, 
superimposed onto the corresponding WSI, in Figure~2.

\subsection*{Removal of puddles of red blood cells}

 This part of the method is rather cumbersome and empirical, but we 
found it useful since, on several WSI of brain tumours, the puddles of 
red blood cells occupied more than 30\% of the regions of blur-free 
tissue. It was developed on two example WSI but proved efficient on all.

 We constructed a mask of large puddles of red blood cells on the WSI at 
objective magnification $40\times$ in the following way. We changed the 
colour space to HSB (as defined in ImageJ) and operated on the B channel. 

 First (step 1), on the result of the contour detection (application of 
a $3\times3$ Sobel filter, ImageJ's \texttt{Find Edges} command), we 
selected connected regions of at least 20 pixels at levels 250 and 
above, and pixels at the distance at most 10 pixels from these regions, 
then retained, from these enlarged connected regions, only those which 
had at least 10,000 pixels.

 Second (step 2), we selected pixels of brightness exactly 255 and 
performed a morphological \texttt{close} operation on the resulting 
mask. This is to select the sides of the red blood cells, which appear 
very bright in bright field microscopy (because of the refringence 
property of these cells) and elongated (because of the circular shape of 
these cells). We skeletonised the result and eliminated the connected 
regions of at most 9 pixels. Then we extended the remaining skeletons to 
pixels within the distance of 20 pixels and eliminated the connected 
regions of less than 20,000 pixels. Defining the threshold $b := 255 - 
0.35 (255 - b_{bg,r})$, were $b_{bg,r}$ was defined during background's 
brightness calibration (see above), we retained, among the remaining 
regions, only those which contained at least one connected regions of 40 
pixels or more with brightness above $b$ and of circularity at most 0.2. 
We also added the connected regions of 2800 pixels and above which were 
constructed by added to these latter regions pixels at distance at most 
2800.

 We took the union of the regions defined in the two preceding steps (1 
and 2). We added the pixels at distance at most 20 pixels of these 
regions and inside the convex hull of at least one of these regions. 
This formed a mask of puddles of red blood cells. Finally, we added to 
the mask the holes of less than 100,000 pixels if contained, and stored 
the mask on the disk as a 1-bit-per-pixel Deflate-compressed tiled TIFF 
image as before.

 A detail from an example of such a mask of large puddles of red blood 
cells is shown, superimposed onto the corresponding WSI, in Figure~3.

 Removing from the mask of sharp tissue pixels considered, according to 
this mask of puddles of red blood cells, we built the mask of 
significant, blur-free tissue.

\subsection*{Calibration of optical density ratios of stains}

 Since the vascular endothelial cells were marked with a brown staining 
over a blue counterstaining, information relevant to the microvessels 
are be entirely contained in the brown channel after we perform a colour 
deconvolution~\cite{colourdeconvolution}.

 To perform this linear change in colour space, we needed to know the 
optical densities (o.d.) in R, G and B channels (or, more precisely, 
only the ratios of these three o.d. to form their vector in the RGB 
space~\cite{colourdeconvolution}) of the two stains, brown and blue. We 
couldn't use standard values from the literature nor common values for 
all WSI since, as for the brightness of the background above, there was 
a strong slide-to-slide variability, as can be seen on Figure~4.

 We used the following procedure on each piece of the $20\times$ mosaic 
without overlap formed earlier. Using ImageJ, we performed a change of 
colour space to HSB. On one hand, we selected pixels with brightness at 
most 198, saturation at least 70 and hue outside the interval 
$[80,199]$. Among them, we kept only connected regions of at least 
25~pixels. Finally, we measured the average optical densities in the R, 
G, and B channels of these pixels and the number of remaining pixels and 
wrote these numbers to a text file. This gave the contribution of 
markedly brown areas. On the other hand, we repeated the procedure with 
pixels, the saturation of which was at least 70 and with hue inside the 
interval $[80, 199]$ to get the contribution of markedly blue areas.

 Then we aggregated the results from all mosaic pieces to compute the 
average over the whole slide of the optical densities in R, G, and B 
channels of the brown and blue stains and wrote them to a text file.

 The precise values of the thresholds above are irrelevant: the 
important thing is that the loose limits on the hue 80 and 199 clearly 
separate brown from blue (they can of course be adapted to other 
colours) and that the loose limits on the saturation and brightness 
select as representative areas for calibration regions markedly brown 
resp. blue. In average, the calibration of the optical densities of the 
brown staining rested on 15,4~megapixels (min: 56,3~kilopixels, max: 
288~megapixels) and the calibration of the optical densities of the 
blue rested on 104~megapixels (min: 1,09~megapixels, max: 
842~megapixels).

\subsection*{Selection of vessel walls}

 The whole process of actual selection of vessel walls operated only on 
the areas of significant, blur-free (sharp) tissue of the WSI, which 
were indicated by the mask constructed earlier.

 First, the histogram $H_B$ of the brown optical density (o.d.) of each 
pixel after colour deconvolution was constructed. This was achieved 
through an in-house developed C program which took as input the WSI at 
resolution~$20\times$, the mask of sharp tissue and the parameters of 
the colour deconvolution (average o.d. determined earlier) and operated 
independently on each tile of the WSI to save RAM and allow the use of 
parallel processing on a computer with multicore CPUs. Here, the 
restriction to areas of sharp tissue also avoided to bias the histogram 
with values from irrelevant pixels (e.g. dust).

 Then, a global (but WSI-specific) threshold on the brown o.d. for 
vessel walls on the WSI, hereafter called $b_A$, was automatically 
determined from the histogram in the following way. First, we computed 
$b_\mathrm{bg}$, the brown o.d. of a pixel with colour $(R_{bg}, G_{bg}, 
B_{bg})$ stored earlier (average colour of the darkest pixels of the 
background reference regions). Then, we applied the \texttt{isodata} 
automatic method of threshold computation~\cite{isodatathreshold} on the 
part of the histogram $H_B$ concerning o.d. above $b_\mathrm{bg}$.

 We could not rely on the full histogram to determine $b_A$ since, on 
some of the WSI, cell nuclei were so dense in the tissue that they would 
manifest themselves as a peak in the low o.d. region of the histogram 
$H_B$, so that the value $b_1$ computed by the \texttt{isodata} 
algorithm would be influenced by them instead on yielding information on 
the vessel walls only. Disregarding the low o.d. values of brown, that 
is, o.d. values below those of the background, was a simple and 
efficient way to solve this problem.

 Then, a second global (but also WSI-dependent) threshold, called $b_B$, 
was computed by a new application of the \texttt{isodata} algorithm on 
the part of the histogram $H_B$ concerning o.d. above $b_\mathrm{A}$.

 The actual segmentation of the vessel walls consisted essentially in 
looking for connected sets of pixels, inside the sharp tissue, which had 
brown o.d. above (or equal to) $b_A$ and which contained at least a 
small regions where brown o.d. was above (or equal to) $b_B$. The second 
condition avoided that pale brown regions be inaccurately recognised as 
vessel walls (this concerned for instance a few isolated CD34-positive 
tumour cells).

 This was performed by treating in turn each rectangular zone of size 
roughly $3840\times3840$ of the WSI at resolution $20\times$. The zone 
was extracted using the \texttt{tifffastcrop} program from the 
LargeTIFFTools~\cite{largetifftools,ndpitools-article} (which can 
extract very quickly a rectangular zone from a (possibly huge) tiled 
TIFF image, twice as fast as the \texttt{extract\_area} command of 
\texttt{vips}). The corresponding zone from the mask of sharp tissue was 
extracted. Then an ImageJ macro selected pixels from the extract of the 
WSI inside the sharp tissue zones according to the mask, performed 
colour deconvolution and created a mask of the brown o.d. of pixels 
which satisfy one of the three conditions: the brown o.d. is above (or 
equal to) $b_A$, the brightness (in HSB colour space) is at most 30, or 
the brightness is at most 40 and the saturation at most 127. The two 
latter conditions were necessary because almost black pixels composing 
some of the vessel walls may have low values of brown o.d. (as already 
noticed for brown staining~\cite{colourdeconvolution}) and may be missed 
by the first condition.

 The mask was post-processed in the following way. Let us call A-regions 
the connected regions of pixels selected so far on the basis of $b_A$ 
(discarding regions of less than 75~pixels). We convoluted the image of 
brown o.d. with a Gaussian kernel of standard deviation 0.5~pixel, then 
kept pixels with intensity $b_B$ and above, then connected regions of at 
least 25~pixels of these pixels. Finally, we kept as vessel walls in the 
current rectangular zone the A-regions containing at least a B-region. 
We recorded their mask on the disk as a PNG file.

 As in an earlier step, we used our C program to merge all masks of 
vessel walls in rectangular zones into a single mask at $20\times$ 
resolution stored in a bilevel, Deflate-compressed tiled TIFF file, 
where vessel walls were black and the background was white.

\subsection*{Review of the segmentation results}

 For each slide, we produced a set of files in DeepZoom format from the 
WSI at resolution $40\times$ using \texttt{vips}. We also produced sets 
of files in DeepZoom format for the mask of sharp tissue and the mask of 
vessel walls. And we produced in DeepZoom format the image of the 
contours of vessel walls, where all pixels are transparent, except the 
pixels which belong to a boundary of a vessel wall (black pixels 
surrounded by at least one white pixel).

 These sets of files were uploaded to a secured web server, along with 
a simple HTML file (automatically generated by a simple Perl script) 
calling the JavaScript OpenSeadragon~\cite{openseadragon} library to 
display the superimposition of the slide and, according to what the user 
selects, of the different masks or contours.

 This allowed a convenient quality control of the segmentation of tissue 
and vessels by the pathologist from his office or a meeting room at the 
Hospital, even though the whole image processing was performed on a 
computer in a physics laboratory. Indeed, all that was needed for this 
visualisation was a standard desktop computer with a JavaScript capable 
web browser. A typical session of quality check is displayed on Figure~6.

\subsection*{Material of our cohort}

 We applied our method to a cohort of 129 human patients suffering from 
brain tumours ranging from WHO grade I to grade IV and in various 
locations: posterior fossa, thalamus and hemispheres. The detailed 
numbers are given in Table~1. Such a variety of tumour types served as a 
challenge to our method (see if it is really robust even without human 
intervention) and was also meant to check how much the microvascular 
density was correlated with tumour grade.

 Each sample was prepared the same way: a 5$\mu$m-thick tissue section 
of formalin-fixed, paraffin-embedded tissue was immunostained with a 
monoclonal mouse anti-human CD34 antibody (QBend-10, 
Dako\textregistered, Agilent Technologies, Santa Clara, California, 
USA). The reaction was carried out in an automated immunohistochemistry 
instrument (Benchmark, Ventana Medical Systems\textregistered, Hoffman 
La Roche, Basel, Switzerland). Patients were excluded if the 
pathological sample was insufficient to perform CD34 immunohistochemical 
analysis.

 The resulting 186 tissue sections mounted on glass slides were 
digitised by a Hamamatsu NanoZoomer at objective resolution $40\times$, 
which produced a NDPI file per slide. The average size of the files was 
748.5~MiB (max: 2.54~GiB), representing in total $\approx$100~GiB of 
compressed data. The average size of the images was 7.03~gigapixels (max: 
16.08~gigapixels).

\section*{Results}

 \subsection*{Quality of the segmentation}

 The quality of the segmentation was reviewed during a collective 
meeting in the hospital (involving pathologist, radiologists and 
physicists). Each slide was displayed thanks to an overhead projector 
connected to a computer, itself using a web browser to retrieve images 
from our web server. Out of the 186 slides, 30 had unfortunately to be 
excluded because they displayed extensive areas of extra-vascular 
CD34-positive cells.

 On a few other slides, the pathologist was able to select zones free 
from extra-vascular CD34 signal. These zones were defined using ImageJ as 
unions of polygons on the $2.5\times$ resolution versions of the WSI and 
stored on the disk as ImageJ's \texttt{.roi} files. The corresponding 
slides were reprocessed by disregarding all pixels outside the selected 
zones for tissue and vessel wall selection.

 On two other slides, it proved sufficient to manually raise slightly 
the threshold $b_A$ and to reprocess the slide (vessel selection step) 
to prevent selecting most extra-vascular CD34-positive cells.

 Otherwise, the segmentation was judged of very good quality by the 
pathologist. An example of the complete segmentation is shown in 
Figure~5.

 \subsection*{Density of microvessels}

 On each WSI, we counted the number of selected pixels on the mask of 
sharp tissue and on the mask of vessel walls, and we took the ratio to 
get the density of microvessels. The measurements were done by an 
in-house C program for the sake of speed and RAM economy, again treating 
images tile-wise to enable multicore parallel processing.

 The distribution (histogram) of the densities is displayed in Figure~7. 
We find that it is remarkably correlated with the tumour grade.

 \subsection*{Computation time performance}

 All processing of the WSI was done on a Mac mini computer with 16~GiB 
of RAM and a quad-core~i7 CPU at 2.3~GHz bought in October 2012. 
Although this computer was rather modest in regard of today's standards, 
we found the overall treatment time to be quite acceptable.

 Figure~8 shows the distribution of the computing time of the vessel 
wall segmentation (including colour deconvolution), which is one of the 
most time-consuming operations. One can see that most WSI could be 
treated in less than 15~minutes. This has to be compared to the time 
necessary to transfer to the web server the OpenSeadragon files of the 
WSI and of the masks: the latter was larger even using a large bandwidth 
network connection.

 Let us also notice that selecting the vessel walls only in the areas of 
sharp tissue of the WSI saved a substantial amount of computation time, 
since the average fraction of the WSI occupied by the sharp tissue was 
only 26.8\% (ranging from 1.9\% to 63.6\%). Most of the rest of the WSI 
was usually background, empty space.

\section*{Discussion}

 \subsection*{Inter-slide variability and robustness}

 As we already discussed, several of the physical parameters of the WSI 
(colour temperature of the background, optical densities of the 
stains,...) vary largely from one slide to the next. This is why our 
method includes several steps of calibration. Some previous 
studies~\cite{angiopath} used values of the parameters of the 
segmentation common to all slides, but this was for smaller sets of 
slides. In our heterogeneous set of 186 slides, this would yield wrong 
results. For instance, the thresholds $b_A$ resp. $b_B$ on brown o.d. 
for vessel segmentation have the following statistics: average 93.2, 
min.~44, max.~160 resp. average 178.5, min.~159, max.~250.

 We believe that other parameters (parameters which were not chosen 
after a calibration) fall in two categories: parameters, the value of 
which is not very relevant, and parameters which might influence the 
final result of the WSI. In the first category fall e.g. the precise 
limits of the interval of hue used to define the reference areas for the 
calibration of brown and blue optical densities. Changing these limits 
will alter only very marginally (if at all) the measured o.d., hence the 
vessel segmentation. It is therefore not worth to give too much 
attention to them.

 Parameters in the second category include e.g. the threshold used to 
fix the limit between sharp and blurred tissue, and most of the 
geometrical parameters of the segmentation of tissue (the minimal / 
maximal number of pixels of connected regions to be kept / disregarded, 
the maximal distance of a small piece of tissue to a large piece to be 
considered as significant,...). We did only a manual calibration of 
them, based on a representative set of a few slides or a few excerpts of 
slides. But, even if a different choice for their value could change the 
final measured microvessel density, we think that this change is less 
important than in the case of calibrated parameters (like stains' o.d.) 
and that it would be systematic, affecting the measure roughly in the 
same way on all WSI. Hence fixing these values doesn't preclude the use 
of our automatic measurement for assistance to tumour grading.

 Finally, let us remark that, even with these possible limitations in 
mind, our method should be much more reproducible than manual 
measurements performed on a few chosen excerpts in each slide even by a 
trained 
pathologist~\cite{choix-resolution-pour-quantification-vaisseaux}, if 
the latter is at all doable on such a large set of slides. We prefer to 
save the pathologist's time for quality control and make our overall 
study cheaper by the extensive use of the computer. Notice also that, 
among our 186 slides, many have very fragmented tissue (all the more 
that we are dealing with fragile brain tissue), making it very difficult 
to estimate accurately the area of tissue. And, of course, a systematic 
manual vessel segmentation is not in order: there is an average of 16571 
vessel fragments in the sharp tissue area of our slides (min.~58, 
max.~72614).

 \subsection*{Impact of red blood cells}

 The fraction of blur-free tissue occupied by puddles of red blood 
cells, as determined by the method above, ranged from 0.04\% to 37.2\%. 
In average, such spreads of red blood cells occupied 4.87\% of the 
blur-free tissue, so they were not a crucial issue for the determination 
of the microvascular density for most slides. But for 8 of our 186 
slides they occupied more than 25\% of the apparent tissue, and for 21 
of them they occupied more than 10\% of the apparent tissue. For these 
slides, had we counted the puddles of red blood cells as tissue, we 
would have underestimated the density of microvessels by 10 to 25\%, 
which is huge in comparison, e.g., to the difference of microvascular 
density between low-grade and high-grade tumours (Figure~7).

 Therefore, we believe that the step of automatic detection of puddles 
of red blood cells, although cumbersome and time-consuming, is 
necessary. We couldn't save time by performing it on the $10\times$ 
magnification WSI instead of the $40\times$ WSI because the specifically 
high intensity pixels of the red blood cells' sides (due to their 
refringence) was lost during the resolution reduction.

 \subsection*{Uncertainties}

 We tried to assess the uncertainties on the measurement of the 
microvessel density against the most important parameters of the method 
in the following way. \begin{itemize} \item Changing by one (over 255) 
the brown o.d. threshold $b_A$ changed by 0.5\% the total area of the 
regions recognised as vessel walls. Therefore, the uncertainty on $b_A$ 
and $b_B$, as determined automatically by a thresholding 
algorithm~\cite{isodatathreshold} up to a few units over 255, has little 
influence on the final result. But remember, once again, that this automatic 
threshold varies strongly from one slide to the next one: the standard 
deviation on $b_A$ is 21.1. \item Eroding or dilating either the regions 
recognised as vessel walls or the regions recognised as tissue by one 
pixel at resolution $20\times$ changed by $\approx$7\% the area. This is 
more serious and means that one has little latitude on morphological 
post-processing of the segmented regions. However, a change by 7\% is 
still acceptable if one wants to use the microvessel density to 
distinguish between low-grade and high-grade tumours, owing to the large 
spreading of the density (see Figure~7). It should be interesting to 
redo the measurement using the WSI at magnification $40\times$ and see 
if the uncertainty against eroding or dilating is smaller as one could 
expect, at the expense of a computation time four times larger. 
\end{itemize}

 In a context of strongly heterogeneous tissue, measuring the 
microvessel density on whole tissue sections also contributes to reduce 
the uncertainties by allowing the measurement to rest on a large zone of 
tumour tissue, hence reducing the risk to measure accidentally a zone 
with especially low or high vessel density.

 Additionally, if one wants to draw a link between measurements at the 
microscopic scale (on WSI) where individual cells are resolved, and at 
macroscopic scale (using e.g. MRI) where the resolution is routinely of 
the order of 1~mm, one has to perform the microvessel density 
measurement on the largest (but still relevant) piece of tissue. In this 
spirit, and to further confirm the quality and relevance of our 
measurement on brain tumour tissues, it has been shown that the 
microvascular density we measured is in good correlation with the result 
of a noninvasive, macroscopic measurement using the ASL modality of 
MRI~\cite{volodia-asl}.

 \subsection*{Further development}

 We plan to extend this work in several directions. First of all, the 
overall process can still be optimised to reduce the treatment time of 
each slide.

 Then, much information is still left unexploited: on one hand, we plan 
to perform morphometry analyses on the segmented vessel 
walls~\cite{caiman, angiopath, sharma2016}. This could serve as a basis 
for a system of computer aided diagnosis of some of the tumours. And 
this would yield precious data to develop a theoretical model of 
angiogenesis in brain tumours, which hopefully could guide treatments in 
the long term, in the spirit of what is being done e.g. for adult 
low-grade gliomas~\cite{radiotherapie}.

 On the other hand, no information about cell nuclei has been exploited 
yet. It should be relatively easy to perform segmentation on e.g. the 
blue channel after colour deconvolution and get quantitative parameters 
in the same way as for vessels: density of nuclei, morphometry...

 And, beyond morphometry on the black-and-white masks resulting from the 
mere segmentation (thresholding) of biological objects, it could also be 
possible to extract more information from the virtual slides by 
continuously varying the threshold defining A-regions. In this way, a 
series of segmentations is built and can be analysed as would be a time 
series~\cite{kayser2016}, revealing more aspects of the disease than a 
static picture taken at a single time point.

 Finally, angiogenesis has been shown to be of significant value for 
diagnosis and prognostic more generally in 
oncology~\cite{vascularite-significative-pour-carcinome-ovarien, 
vascularite-barrett, Szoke01052007}, so that our method can readily be 
applied to other tumours with the same of similar immunostainings (CD31, 
CD34).

\section*{Conclusions}

 We have introduced an automatic and training-free method of 
quantification of the density of microvessels on whole tissue sections 
immunostained with the CD34 antibody and digitised by a slide scanner. 
This method is, to our knowledge, the first one to include a careful 
determination of areas of tissue without blur and puddles of red blood 
cells before the proper segmentation of vessel walls.

 We tested in on a large set of WSI (186) of a very large variety of 
brain tumours. Using a very reasonable amount of computation time on a 
quite affordable computer system (an Intel Core i7 CPU with 16~MiB of 
RAM), this method produced results of very good quality, even though an 
overall check of the segmented WSI by the pathologist was necessary, in 
particular because of extra-vascular CD34-positive tumour cells. It 
should be helpful in computer-aided diagnosis systems and easily reused 
for other stainings/tumours, especially because it uses only open source 
software (like ImageJ or vips) or well-described algorithms, and because 
its architecture is simple and modular and its parameters easy to 
understand and modify (e.g. to adapt it to other colours than brown and 
blue).

  \ifthenelse{\boolean{publ}}{\small}{}

\section*{Competing interests}

The authors declare that they have no competing interests.
    
\section*{Acknowledgements}

We thank the CNRS/IN2P3 computing centre for hosting the website we used 
to view WSI and control the quality of segmentation.

CD and MB belong to the CNRS consortium CellTiss and to the Labex P2IO.
 

{\ifthenelse{\boolean{publ}}{\footnotesize}{\small}
 \bibliographystyle{bmc_article}  
  \bibliography{vascul_volodia} }     


\ifthenelse{\boolean{publ}}{\end{multicols}}{}



\section*{Figures}

\begin{minipage}{\linewidth}
\begin{center}

\begin{tikzpicture}
 [show background rectangle, background rectangle/.style={fill=gray!20},
  image/.style={rectangle,draw,fill=blue!30},
  parametres/.style={rectangle,rounded corners=0.75ex,draw,fill=red!40},
  mesure/.style={rectangle,rounded corners=0.75ex,draw,fill=green!70!black!30},
  masque/.style={rectangle,draw,pattern=checkerboard,pattern color=white},
  f/.style={-triangle 45,thick},
  invisible/.style={inner sep=0,minimum size=0}]
\node [image] (40xWSI) {$40\times$ WSI};
\node [image,below=of 40xWSI] (2p5xWSI) {$2.5\times$ WSI};
\node [parametres,below=of 2p5xWSI] (minandmaxbackgrbr)
  {\parbox{3.3cm}{\centering min and max values\\of background's\\
    brightness}};
\node [masque,right=of minandmaxbackgrbr] (maskoftissue) {mask of tissue};
\node [invisible,above=6mm of maskoftissue] (inputmaskoftissue) {};
\path (minandmaxbackgrbr.east) -- (maskoftissue.west)
  node[invisible,midway] (minmaxbackgrtoinputmaskoftissue) {};
\node [image] (20xmosaicwithov) at (2p5xWSI -| maskoftissue)
  {\parbox{2cm}{\centering $20\times$ mosaic\\ with overlap}};
\node [masque,right=of maskoftissue] (maskofsharp)
  {\parbox{2cm}{\centering mask of\\sharp areas}};
\node [image,above=of maskofsharp] (20xmosaicwithoutov)
  {\parbox{2cm}{\centering $20\times$ mosaic\\ w/o overlap}};
\node [image] (20xWSI) at (40xWSI -| 20xmosaicwithov) {$20\times$ WSI};
\node [masque,left=of minandmaxbackgrbr] (maskofrbc)
  {\parbox{3cm}{\centering mask of puddles\\of red blood cells}};
\node [invisible,below=of maskoftissue] (inputmaskofsharptissue) {};
\node [masque,below=of inputmaskofsharptissue] (maskofsharptissue)
  {\parbox{2cm}{\centering mask of\\sharp tissue\\w.o. puddles\\of r.b.c.}};
\node [mesure,right=of maskofsharptissue] (area)
  {\parbox{2cm}{\centering area of\\tissue}};

\draw[f] (40xWSI.east) to (20xWSI.west);
\draw[f] (40xWSI.south) to (2p5xWSI.north);
\draw[f] (20xWSI.south) to (20xmosaicwithov.north);
\draw[f] (20xWSI.east) -| (20xmosaicwithoutov.north);
\draw[f] (2p5xWSI.south) to (minandmaxbackgrbr.north);
\draw[f] (20xmosaicwithoutov.south) to (maskofsharp.north);
\draw[f] (40xWSI.west) -| (maskofrbc.north);

\draw[thick] (minandmaxbackgrbr.east) to (minmaxbackgrtoinputmaskoftissue.east);
\draw[thick] (minmaxbackgrtoinputmaskoftissue.south) |- (inputmaskoftissue);
\draw[f] (20xmosaicwithov) to (maskoftissue);

\draw[thick] (maskofrbc.south) |- (inputmaskofsharptissue);
\draw[thick] (maskofsharp.south) |- (inputmaskofsharptissue);
\draw[f] (maskoftissue.south) to (maskofsharptissue);

\draw[f] (maskofsharptissue.east) to (area.west);
\end{tikzpicture}

\bigskip

\begin{tikzpicture}
 [show background rectangle, background rectangle/.style={fill=brown!20},
  image/.style={rectangle,draw,fill=blue!30},
  parametres/.style={rectangle,rounded corners=0.75ex,draw,fill=red!40},
  mesure/.style={rectangle,rounded corners=0.75ex,draw,fill=green!70!black!30},
  masque/.style={rectangle,draw,pattern=checkerboard,pattern color=white},
  f/.style={-triangle 45,thick},
  invisible/.style={inner sep=0,minimum size=0}]
\node [image] (20xWSI) {\parbox{2.5cm}{\centering $20\times$ WSI (only\\
  zones selected\\using the mask\\of sharp tissue\\are considered)}};
\node [parametres,right=of 20xWSI] (odratios)
  {\parbox{3cm}{\centering o.d. ratios of\\the brown and\\blue stains}};
\node [invisible, below=of odratios] {};
\node [parametres,right=of odratios] (histogram)
  {\parbox{3cm}{\centering histogram of the\\brown o.d.}};
\node [image,fill=brown!40,below=of histogram] (brown)
  {\parbox{2.5cm}{\centering image of brown\\o.d. after\\colour deconv.}};
\node [parametres,right=of histogram] (thresholds)
  {\parbox{2.8cm}{\centering two automatic\\thresholds\\on the brown o.d.}};
\node[masque] (mask) at (thresholds.south |- brown.east)
  {\parbox{2.8cm}{\centering mask of vessel\\walls inside\\sharp tissue}};
\node [mesure,below=of mask] (area) {area of vessel walls};

\path (brown.east) -- (mask.west) node[invisible,midway] (inputmask) {};
\path (thresholds.south) -- (mask.north) node[invisible,midway]
  (thresholdstoinputmask) {};

\draw[f] (20xWSI.east) to (odratios.west);
\draw[f] (20xWSI.south) |- (brown.west);
\draw[f] (odratios.south) |- (brown.west);
\draw[f] (brown.north) to (histogram.south);
\draw[f] (histogram.east) to (thresholds.west);

\draw[thick] (thresholds.south) to (thresholdstoinputmask.south);
\draw[thick] (thresholdstoinputmask.east) -| (inputmask);
\draw[f] (brown.east) to (mask.west);

\draw[f] (mask.south) to (area.north);

\end{tikzpicture}

\end{center}
  \subsection*{Figure 1 - Overview of the whole method of selection of 
   sharp tissue (top) and of vessel walls inside sharp tissue (bottom)}

The blue boxes represent real, full colour images; the checkerboard boxes 
represent masks (black-on-white images resulting from the selection of 
objects); the red boxes represent automatically determined quantitative 
parameters; the brown box represents the image of the brown optical 
densities of pixels of the original $20\times$ image; the green boxes 
represent the final measured quantities.

\end{minipage}

\begin{minipage}{\linewidth}
\begin{center}
\includegraphics[width=0.49\linewidth]{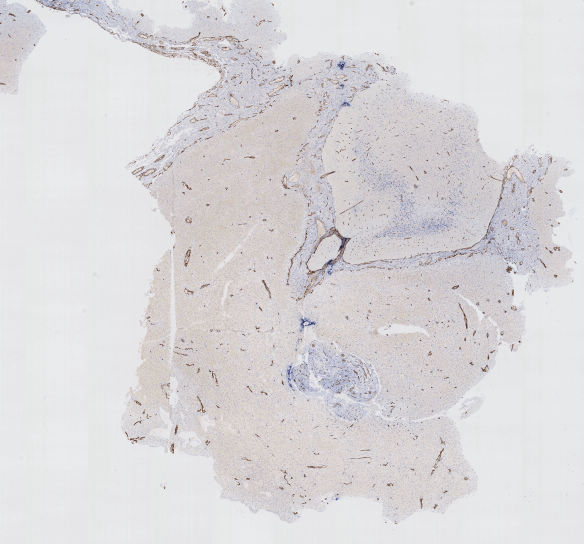}
\hfill
\includegraphics[width=0.49\linewidth]
 {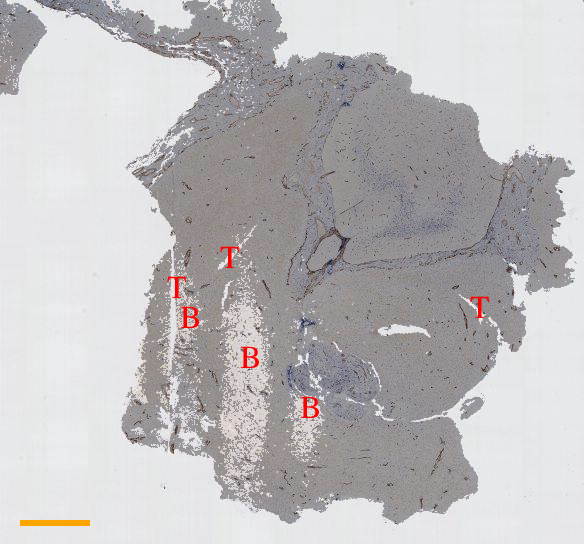}
\par \medskip
\includegraphics[width=0.49\linewidth]{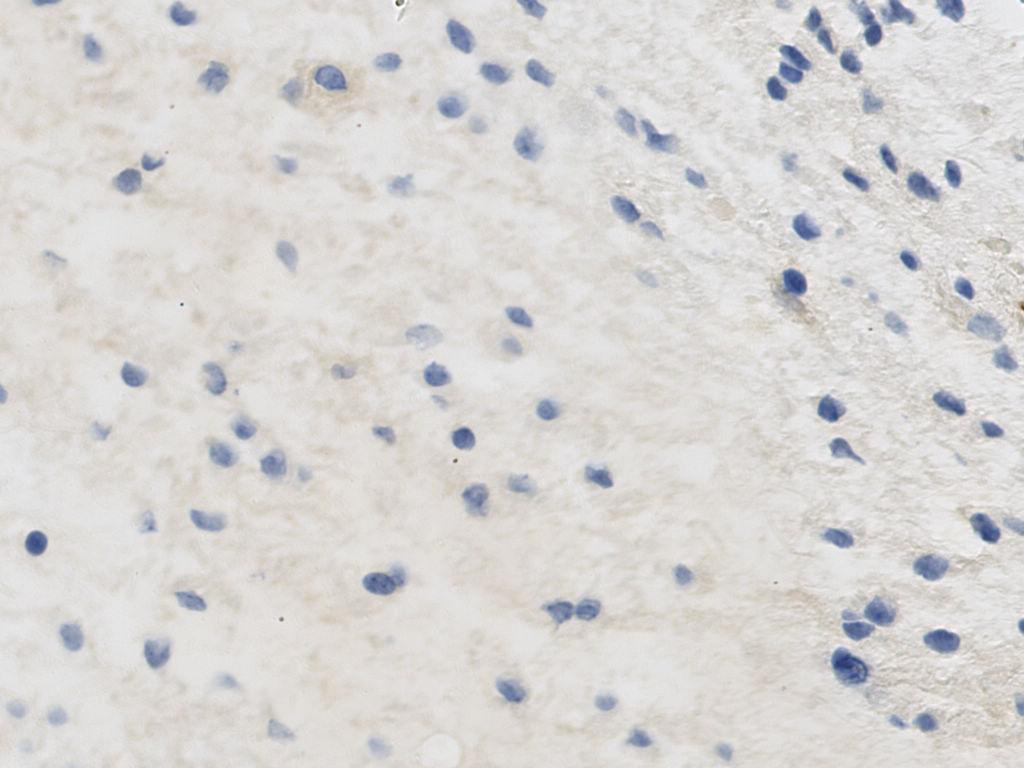}
\hfill
\includegraphics[width=0.49\linewidth]
 {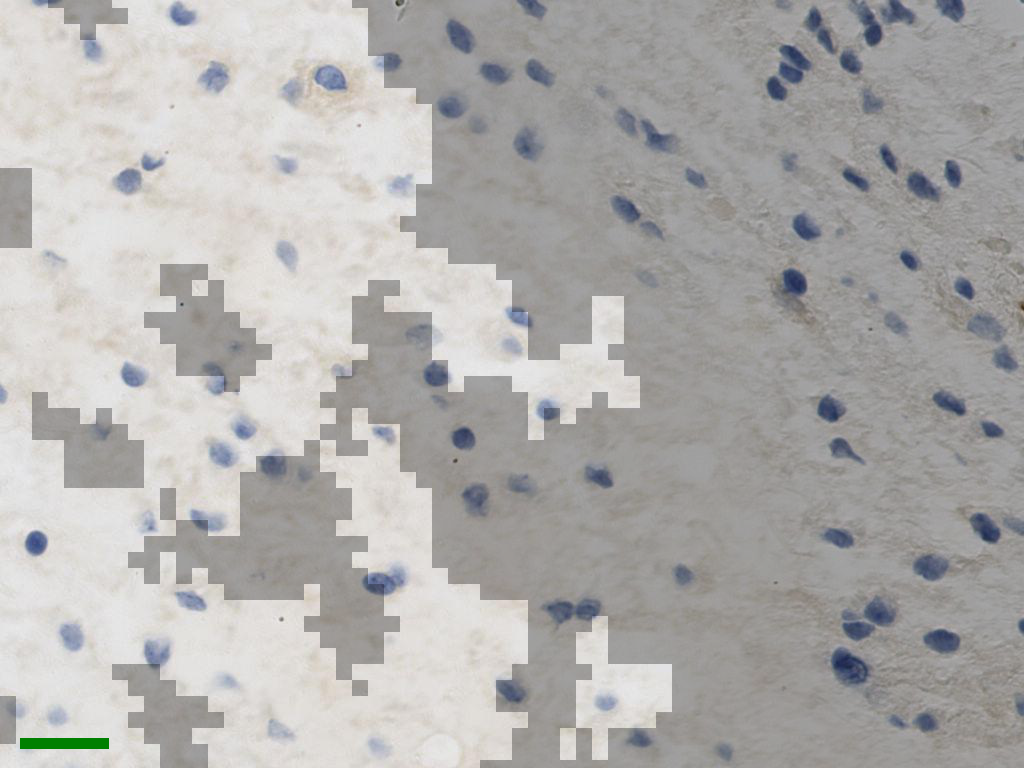}

\end{center}
  \subsection*{Figure 2 - Example of a determination of tissue without 
  blur.}

{\bf Top left:} excerpt of the original image (one of our WSI). {\bf Top 
right:} the mask (logical \texttt{and} combination of the mask of tissue 
and of the mask of sharp regions) is superimposed onto the WSI. Only 
regions considered, after our method, as sharp tissue are shaded. One 
can see in particular that tears inside the tissue are properly not 
counted as tissue (some of them are marked `T'). Blurred regions (some 
of them are marked 'B') tend to form three vertical bands because of the 
way the image was acquired by the scanner. They will be excluded from 
the quantification process. Orange scale bar = 1~mm. {\bf Bottom left:} 
detail of the top right image at the boundary of a blurred region. {\bf 
Bottom right:} the mask of sharp regions is superimposed onto the detail 
of the WSI shown on the bottom left. One sees the transition between 
``fully blurred'', on the left-hand side of the image, to ``fully 
sharp'', on the right-hand side of the image. Green scale bar = 
20~microns.

\end{minipage}

\begin{minipage}{\linewidth}
\begin{center}
\includegraphics[width=\linewidth]{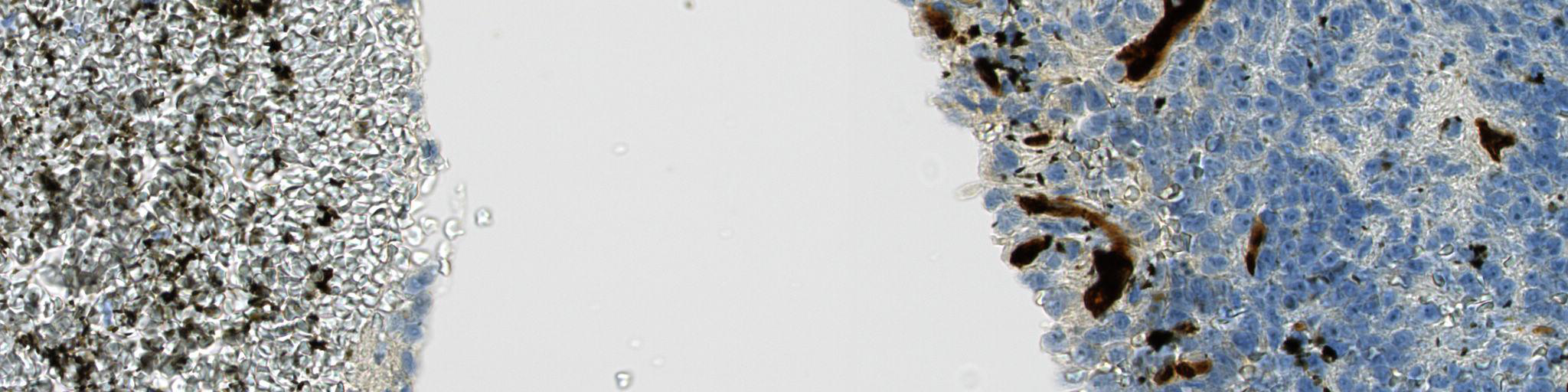}
\par \smallskip
\includegraphics[width=\linewidth]{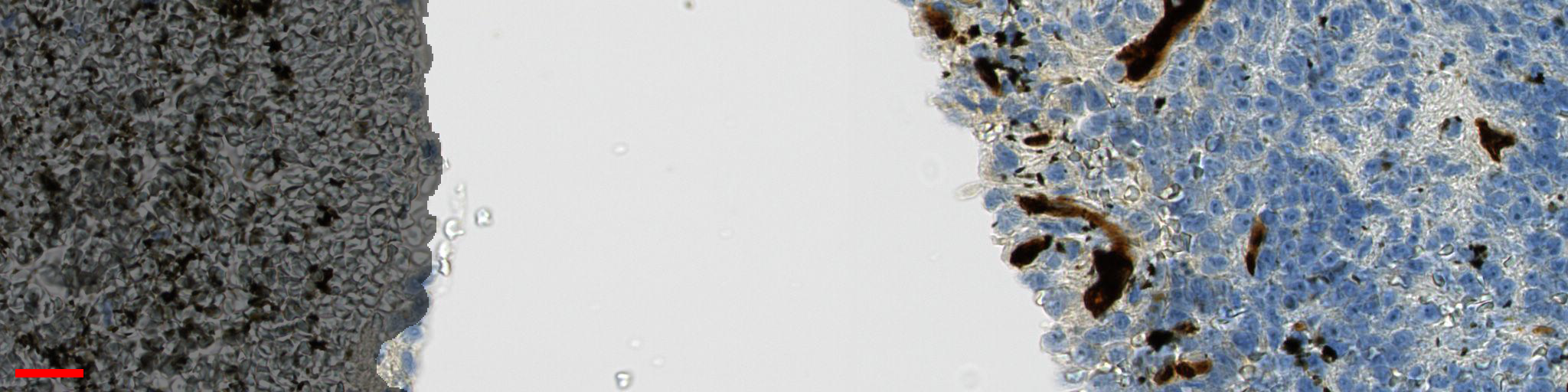}
\end{center}
  \subsection*{Figure 3 - Example of a determination of large puddles of 
  red blood cells.}

{\bf Top:} original image (excerpt of one of our WSI). {\bf Bottom:} the 
same image, over which the mask of the large puddles of red blood cells 
was superimposed. Scale bar = 20~microns.

\end{minipage}

\begin{minipage}{\linewidth}
\begin{center}
\includegraphics[width=\linewidth]{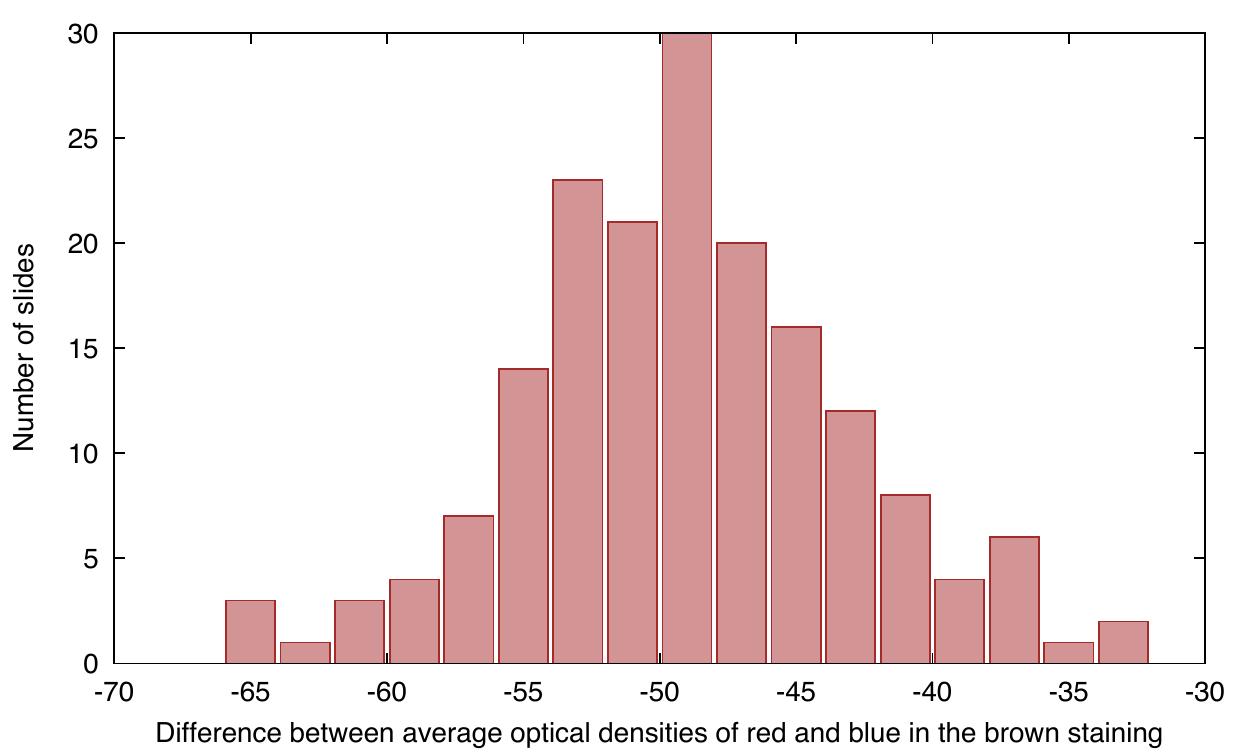}
\end{center}
  \subsection*{Figure 4 - Slide-to-slide variability of the optical 
densities of the brown staining.}
 For each WSI, after calibrating the optical densities in the red, green 
and blue channels of the brown (CD34) staining as defined 
in~\cite{colourdeconvolution} according to the procedure described in 
the main text, we subtracted the blue optical density to the red optical 
density --- let's call this difference $\Delta$. This histogram shows 
the distribution of the values of $\Delta$ over the 186 WSI. One can see 
that there is a strong variability, which prevents one to use a single 
set of optical densities to perform colour deconvolution on all WSI. 
\end{minipage}

\begin{minipage}{\linewidth}
\begin{center}
\includegraphics[width=\linewidth]{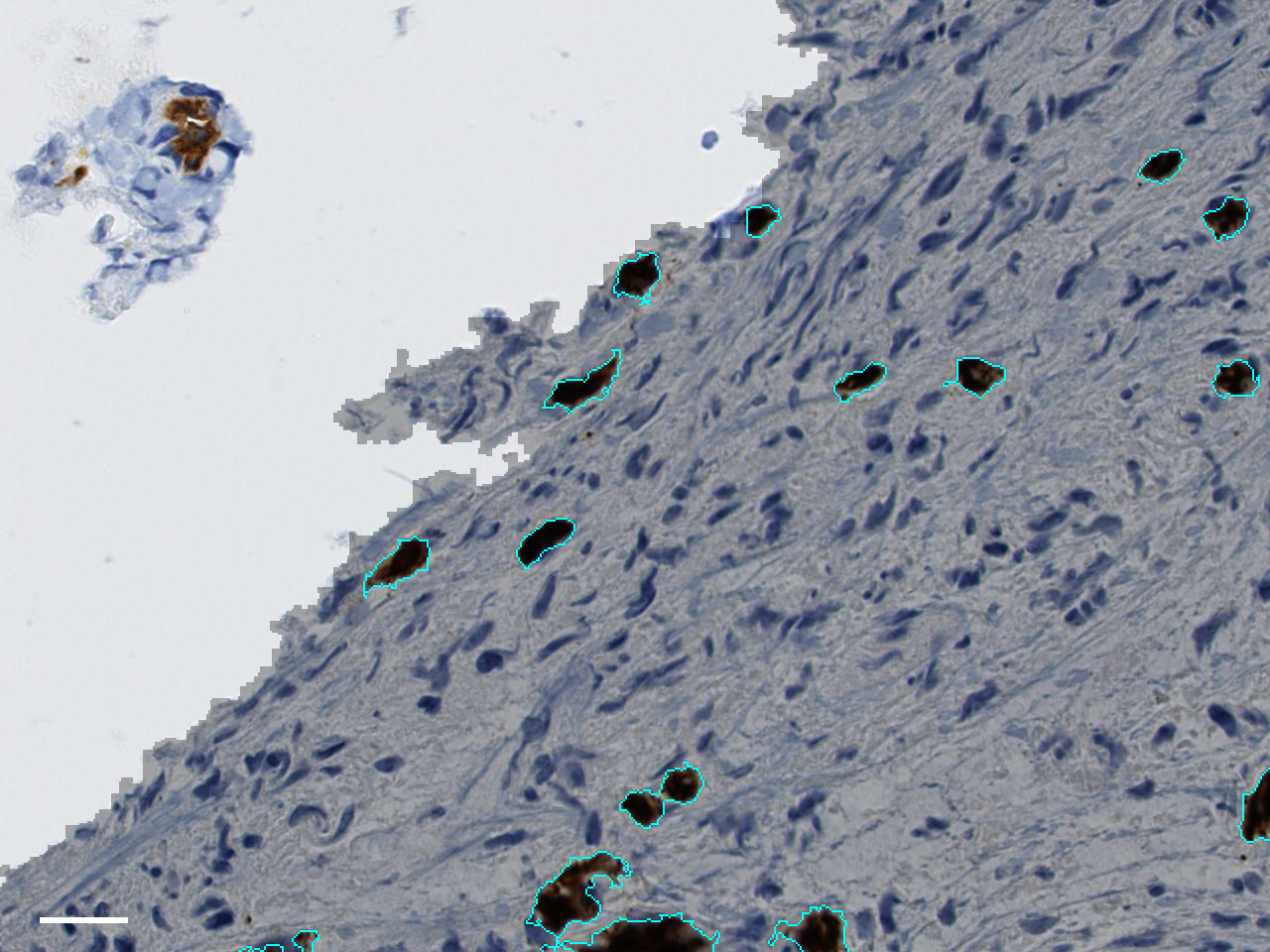}
\end{center}
  \subsection*{Figure 5 - Example of a segmentation of tissue and vessel 
walls (detail).}
On this excerpt of a WSI at resolution $40\times$, the final result of 
the segmentation by our method is shown. The areas considered as sharp 
tissue are shaded, and the areas inside sharp tissue considered as 
vessel walls on the basis of the CD34 immunostaining are contoured in 
cyan. Scale bar = 20 microns.
\end{minipage}

\mbox{}\bigskip\mbox{}

\noindent \begin{minipage}{\linewidth}
\begin{center}
\includegraphics[width=\linewidth]{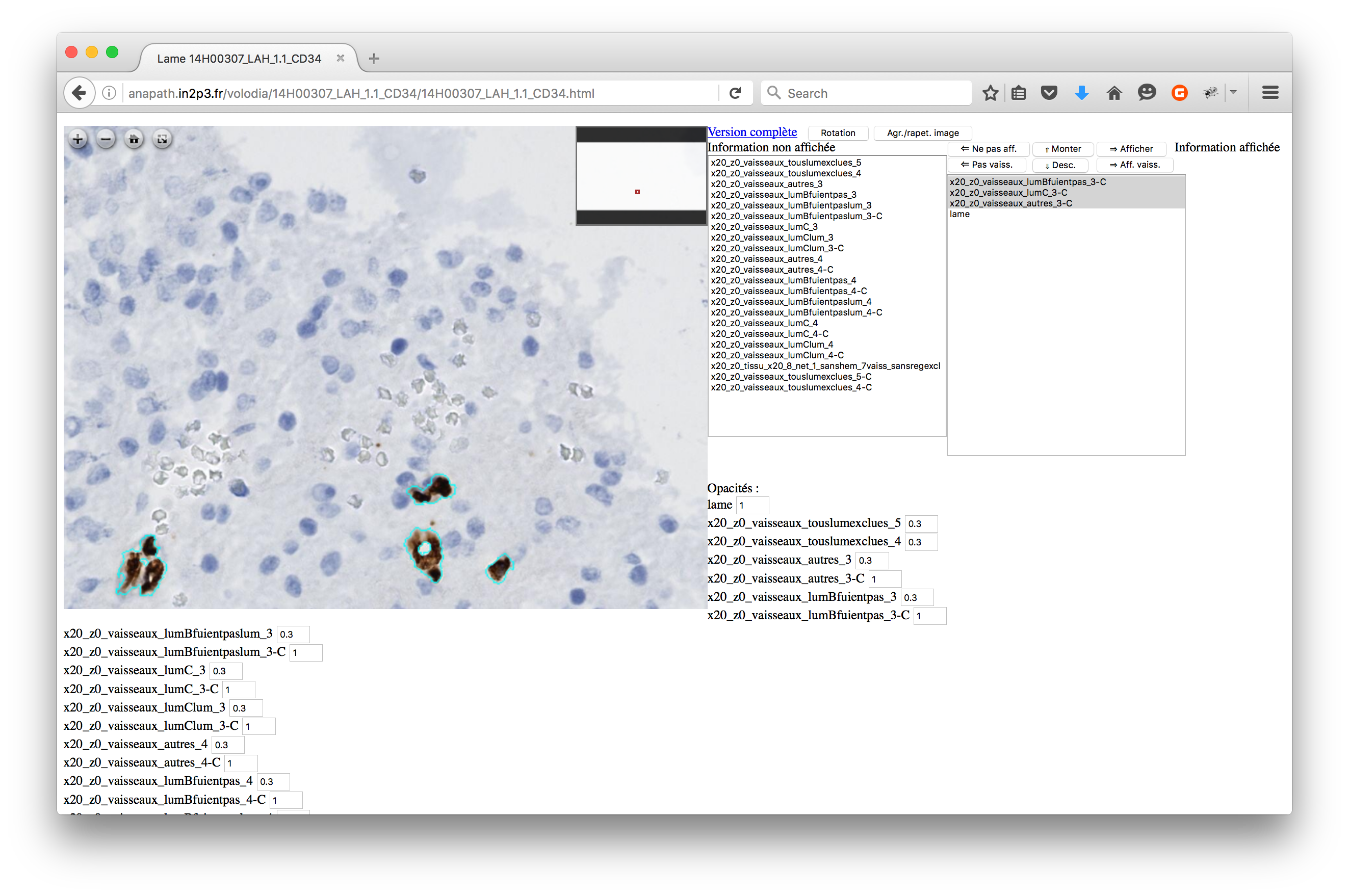}
\end{center}

  \subsection*{Figure 6 - A typical session of quality control of the 
  segmentation in a standard web browser}
 The user is viewing, in the window of his JavaScript-capable web browser 
(here, Firefox), a detail of one of the whole slide images at full 
resolution ($40\times$), on which he has superimposed the contours of 
the vessel walls (displayed in cyan on top of the slide). He can 
interactively add/remove contours and masks (displayed by shading the 
slide) to/from the list of displayed information, zoom in and out and 
drag the slide to explore it with his mouse.
\end{minipage}

\noindent \begin{minipage}{\linewidth}
\begin{center}
\includegraphics[width=\linewidth]{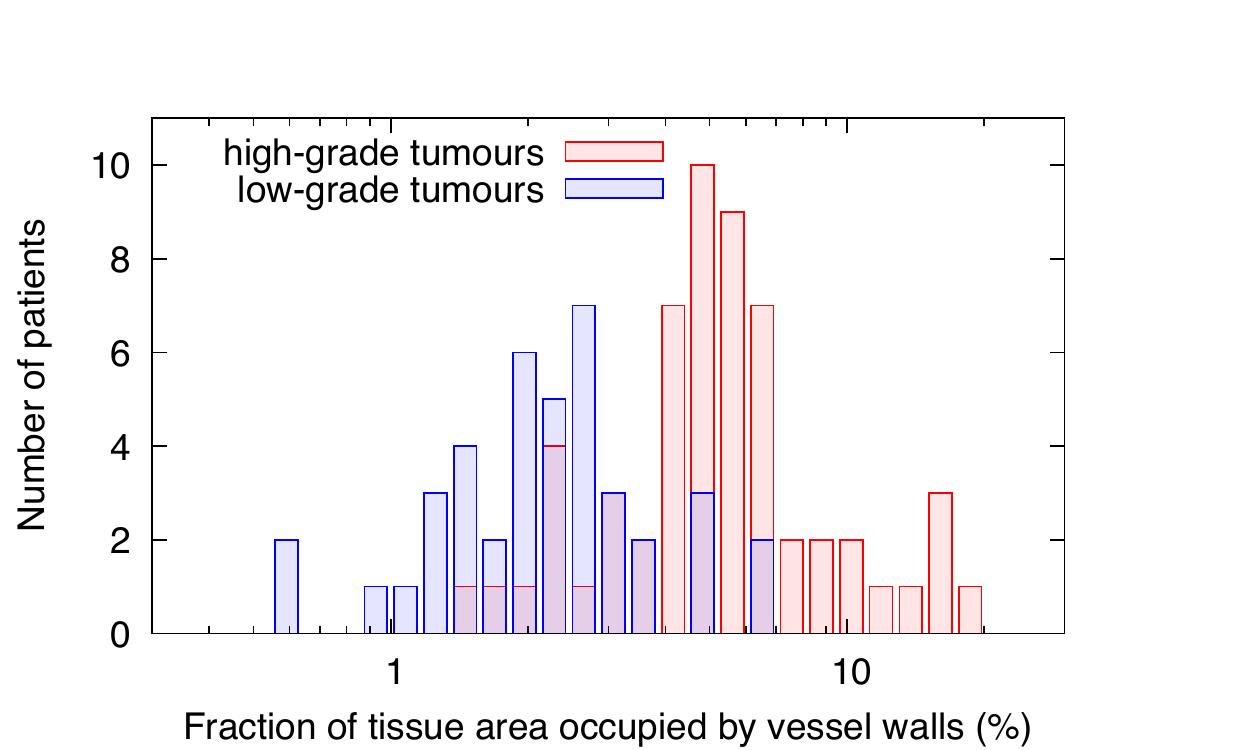}
\end{center}

  \subsection*{Figure 7 - Distribution of the density of microvessels in 
  the tissue for patients suffering a high-grade resp. low-grade 
  tumour}
 The red (resp. blue) histogram shows the distribution of the 
microvascular density (fraction of sharp tissue area covered with 
CD34-stained vessel walls) measured by our method on the WSI of samples 
from high-grade (resp. low-grade) tumours. Although the two histograms 
(here shown in log-lin scale) are relatively broad, there is a clear 
distinction between the typical microvascular densities of low- and 
high-grade tumours. We argue that the measurement uncertainties of the 
microvascular density (see Discussion in the main text) are much lower 
than the difference between these typical values.
\end{minipage}

\noindent \begin{minipage}{\linewidth}
\begin{center}
\includegraphics[width=\linewidth]{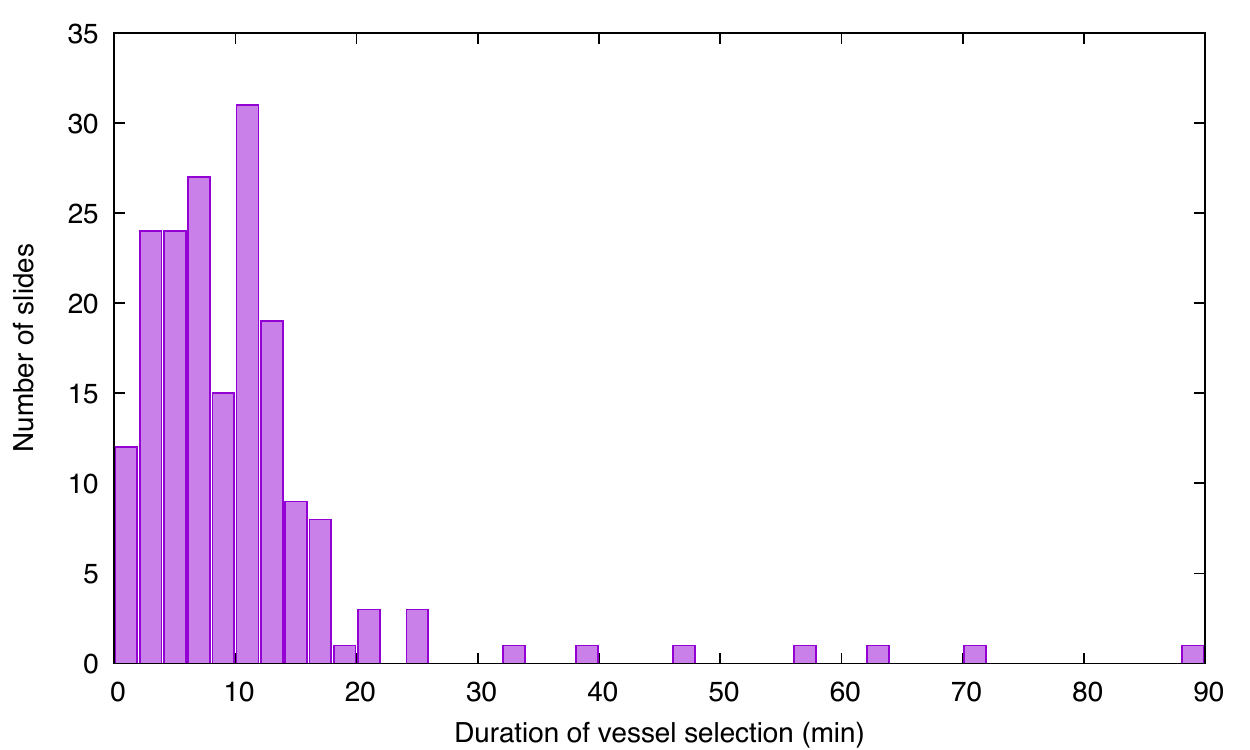}
\end{center}

  \subsection*{Figure 8 - Distribution of the duration of the process of 
colour deconvolution and vessel wall selection of all WSI on our Mac 
mini computer}
 Although a few WSI can request anomalously large treatment times, a 
vast majority of them are treated in less than 15~min on the rather 
modest computer we used for this study.
\end{minipage}

\mbox{}\bigskip\mbox{}


\section*{Tables}
  \subsection*{Table 1 - Types and locations of the 129 paediatric brain 
tumours in our study}
    \par
    \mbox{
\begin{tabular}{|l|r|l|r|l|r|}
\hline
\textbf{Posterior fossa} & \textbf{85} & \textbf{Thalamus} & \textbf{10} &
 \textbf{Hemispheres} & \textbf{34} \\
\hline
Pilocytic astrocytoma & 29 & Pilocytic astrocytoma & 5 &
 Atypical teratoid rhabdoid tumour & 5 \\
Grade I ganglioglioma & 4 & & & Grade I ganglioglioma & 13 \\
Medulloblastoma & 33 & Grade III astrocytoma & 3 &
 Pleomorphic xanthoastrocytoma & 1 \\
Grade III ependymoma & 9 & & & Grade III ependymoma & 3 \\
Grade III ganglioglioma & 1 & & & Grade III ganglioglioma & 3 \\
Other embryonal tumors & 7 & Glioblastoma & 2 & Glioblastoma & 7 \\
Glioma not otherwise spec.  & 1 & & & Glioma not otherwise specified & 1 \\
Hemangioblastoma & 1 & & & Pleomorphic xanthoastrocytoma & 1 \\
 & & & & with anaplastic features & \\
\hline
\end{tabular}
      }

\end{bmcformat}
\end{document}